%% file: CSQCD3.tex
\documentclass[12pt]{article}
\usepackage{epsfig}

\newcommand{\bea}{\begin{eqnarray}}
\newcommand{\eea}{\end{eqnarray}}
\newcommand{\beq}{\begin{equation}}
\newcommand{\eeq}{\end{equation}}
\newcommand{\bce}{\begin{center}}
\newcommand{\ece}{\end{center}}
\newcommand{\ket}[1]{| {#1} \rangle}
\newcommand{\bra}[1]{\langle {#1} |}

\input econfmacros-csqcd3.tex
\def\Title#1{\begin{center} {\Large {\bf #1} } \end{center}}

\begin{document}

\Title{Neutron Star Matter}

\bigskip\bigskip


\begin{raggedright}

{\it Jochen Wambach\index{JW}\\
Institut f\"ur Kernphysik\\
Technische Universit\"at Darmstadt\\
Schlossgarten Str. 2\
64289 Darmstadt\\
Germany\\
{\tt Email: wambach@physik.tu-darmstadt.de}}
\bigskip\bigskip
\end{raggedright}

\section{Introduction}

Neutron stars are the densest  stars in the universe. They are the remnants of violent explosions of massive progenitors in type II supernovae after collapse of the iron core. Neutron-star matter is bound by gravity and the central density can reach several times that in interior of a heavy nucleus. The largest masses currently known are those of the pulsars PSR J1614-2230 and PSR J0348+043 with $1.97\pm 0.04$~\cite{Demorest:2010} and $2.01\pm 0.04$~\cite{Antoniadis:2013} solar masses respectively. Both are in binaries with a white dwarf companion which allows precise measurements of the pulsar mass.

\begin{figure}[hc]
\bce
\includegraphics[width=10.5cm]{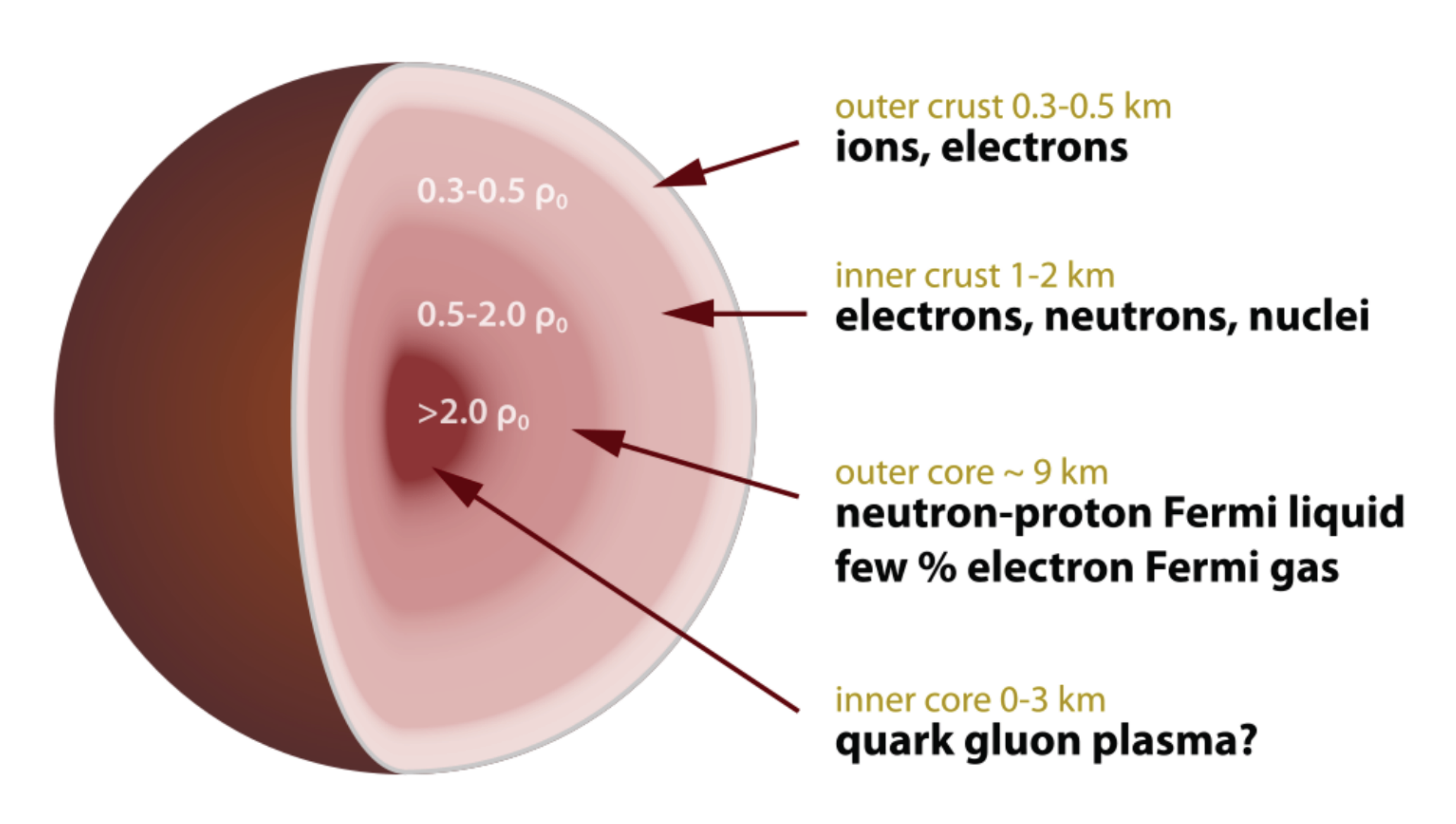}
\ece
\caption{Schematic view of the composition of a neutron star~\cite{Wiki}}
\label{fig:Nstar}
\end{figure}

In General Relativity, the maximum mass of a neutron star is determined by the equation of state (EoS), $P(\epsilon)$, which relates pressure and energy density. The EoS is determined by the composition of the neutron star. A schematic sketch is given in Fig.~\ref{fig:Nstar}. 
\bce
\ece
While the properties of the outer crust and parts of the inner crust are fairly well understood there remain a varity of interesting questions for the deeper interior. These include the possible existence of exotic nuclear shapes (pasta phases) at the interface of the inner crust and the outer liquid core. Also the nuclear pairing properties in the neutron-proton liquid and their relation to the (neutrino) cooling rates remain under debate. One of the most interesting questions, however, relates to the very dense inner core. Since the density can potentially reach values where nucleons start to overlap, one might expect a transition to deconfined quark matter. Whether such a new state of hadronic matter is realized, depends on many details of the high-density EoS, which are poorly understood at present. Here, the high-mass pulsars PSR J1614-2230 and PSRJ0348+043 put severe constraints.

In the following, I will discuss two facets of the high-density EoS. The first relates to properties of the outer liquid core and the role of the symmetry energy. The symmetry energy is one of the terms in the Bethe-Weizs\"acker mass formula for a nucleus of mass number $A$ with $Z$ protons and $N$ neutrons:
\begin{equation}
E_{sym}=a_{sym}\frac{(N-Z)^2}{A};\qquad A=Z+N
\end{equation}
and determines the change in nuclear binding energy with proton-neutron asymmetry. For homogeneous nuclear matter with number densities $n_n$ and  $n_p$ for neutrons and protons, the corresponding symmetry energy, $S(n)$, specifies the difference in energy compared to the symmetric case ($n_p=n_n)$ as
\begin{equation}
\Delta E\simeq S(n)\left(\frac{n_n-n_p}{n}\right)^2;\quad n=n_n+n_p\; .
\end{equation}
As will be detailed below, its density dependence is decisive for the mass-radius relation of neutron stars~\cite{LaPr:2001} and precise constraints from nuclear physics are highly desirable.  

The second part of the discussion deals with a more speculative issue relating to deconfined quark matter in the inner core of a neutron star. Quantum Chromodynamics (QCD) predicts that the (approximate) chiral symmetry of left- and right-handed quarks is spontaneously broken in the vacuum of the strong interaction. This gives rise to a non-vanishing ground-state expectation value $\bra{0}\bar q q\ket{0}$ of the scalar quark-field bilinear which is called the chiral condensate ($CC$). Physically, it is a measure for the non-perturbative generation of a 'constituent' quark mass of around 300-500 MeV. Due to asymptotic freedom, it is expected that the $CC$ vanishes at high density and chiral symmetry is restored. The quarks loose their consttuent mass and aquire their much smaller 'Higgs' masses. As will be discussed, the transtion to the restored phase is likely to proceed through a series of spatially inhomogeneous phases with modulations in the quark density. This situation is similar to the pasta phases at the crust-core boundary. If such phases were to occur in the inner core of a neutron star they could have interesting consequences for transport properties and the interaction with the star's magnetic field.

\section{A new constraint on the Symmetry Energy}

\begin{figure}[hc]
\bce
\includegraphics[width=5.5cm]{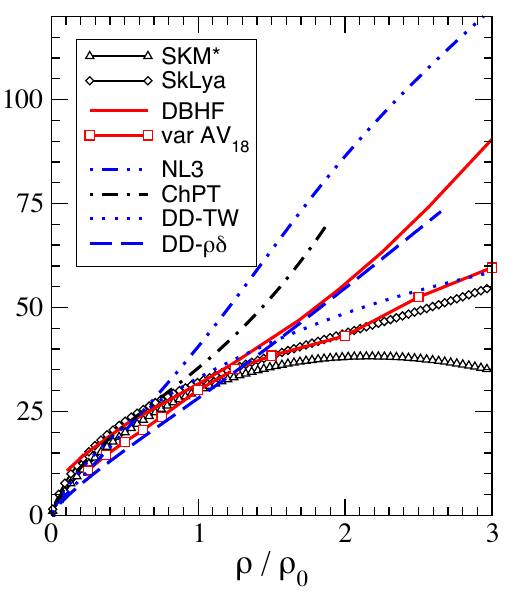}\hspace{0.5cm}\includegraphics[width=7.7cm]{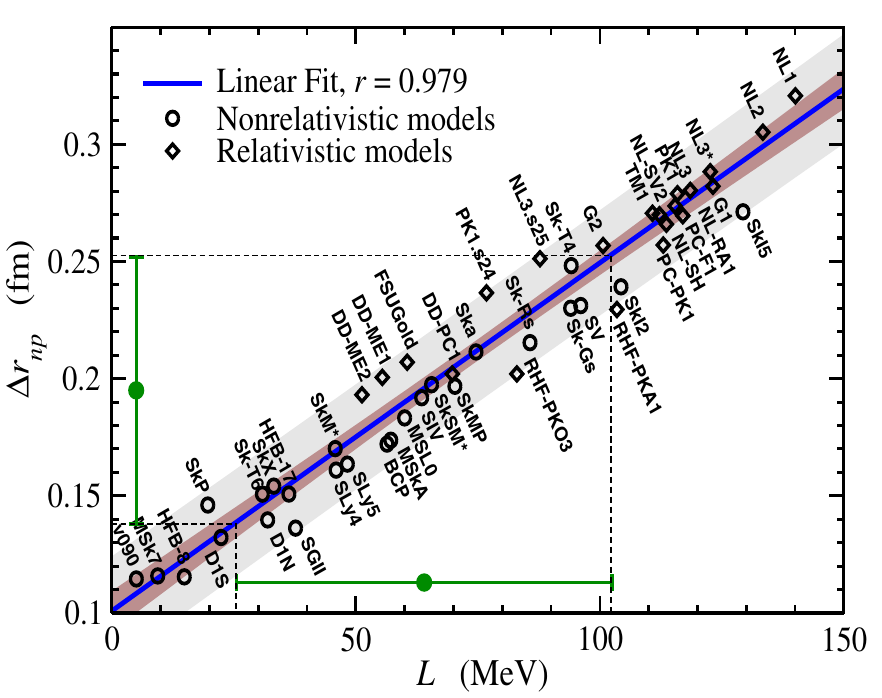}
\ece
\caption{\label{fig:Symm}
Left panel: Density dependence of the nuclear symmetry energy as extrapolated from the properties of known nuclei~\cite{FuWo:2006} ($\rho=n$). Right panel: Correlation between the skin thickness $\Delta r_{np}$ of $^{208}$Pb and the slope $L$ of the symmetry energy at nuclear saturation density, $n_0$~\cite{ACVW:2011}.} 
\end{figure}

Uncertainties in the high-density EoS of neutron-star matter are encoded in the density dependence of the symmetry energy, $S(n)$. A compilation of theoretical extrapolations from the properties of known nuclei is shown in the left panel of Fig.~\ref{fig:Symm} which reveals large uncertainties above nuclear 
saturation density, $n_0=0.16$ fm$^{-3}$.

Of special importance is the slope of $S(n)$ at saturation density: $L=3n_0(dS/dn)_{n_0}$, which is directly related to the pressure $P_0$ of pure neutron matter at this density since $L=3P_0/n_0$~\cite{PiCe:2009} and hence the neutron-star radius. While $L$ is not directly observable, one can use its strong correlation with the neutron skin thickness $\Delta r_{np}$ of a heavy nucleus to obtain experimental constraints. This correlation is displayed in the right panel of Fig.~\ref{fig:Symm} as predicted both in non-relativistic- as well relativistic mean-field models~\cite{ACVW:2011}. The precise measurement of $\Delta r_{np}$ of $^{208}$Pb is the objective of the PREX experiment at the Jefferson Laboratory by using parity-violating electron scattering. The current value of $\Delta r_{np}= 0.34^{+0.15}_{-0.17}$~fm~\cite{AAAAA:2012} still suffers from limited statistics and is to be improved in the future.

\begin{figure}[tbh]
\bce
\includegraphics[width=7cm]{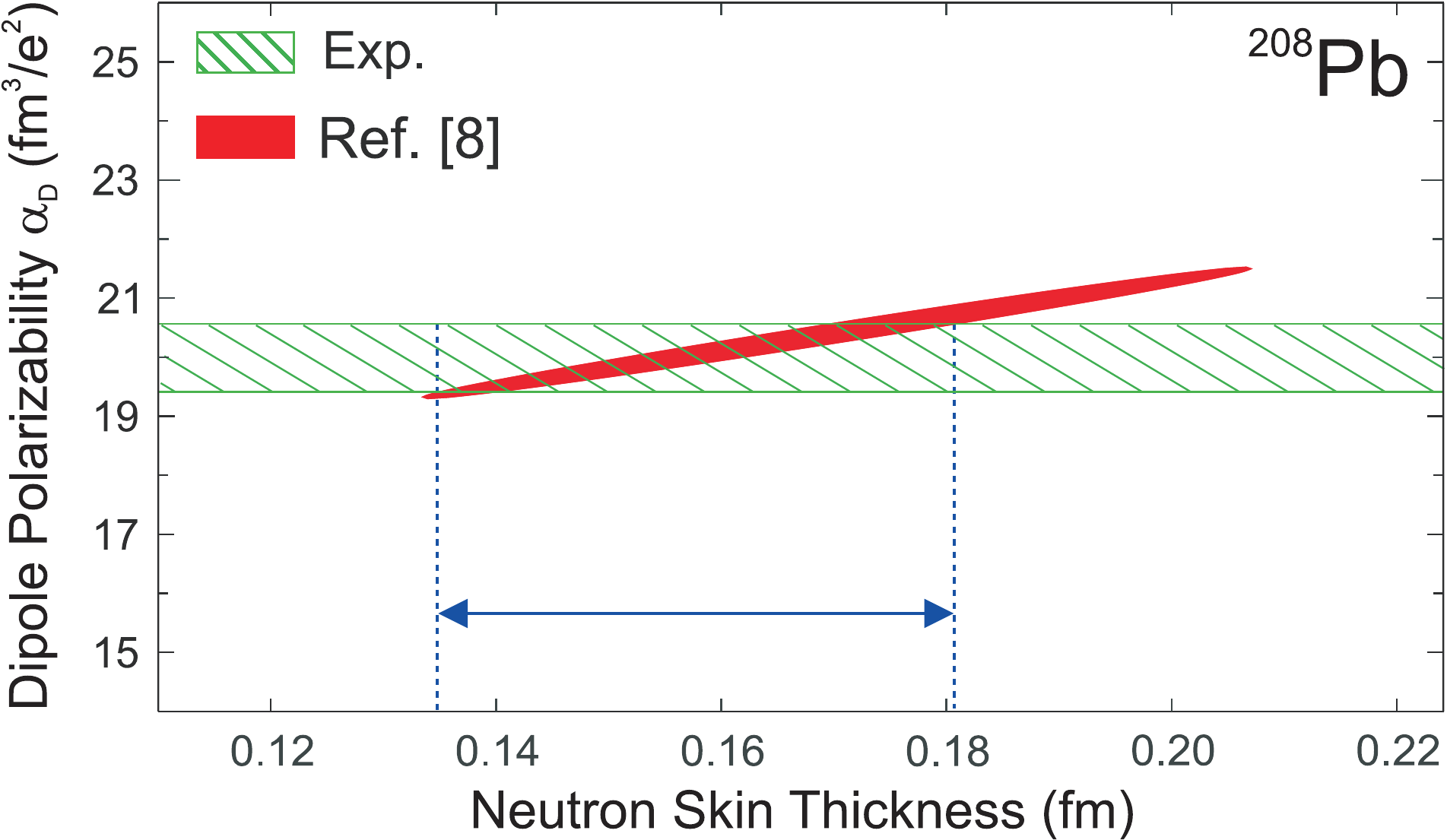}\hspace{0.5cm}\includegraphics[width=5.5cm]{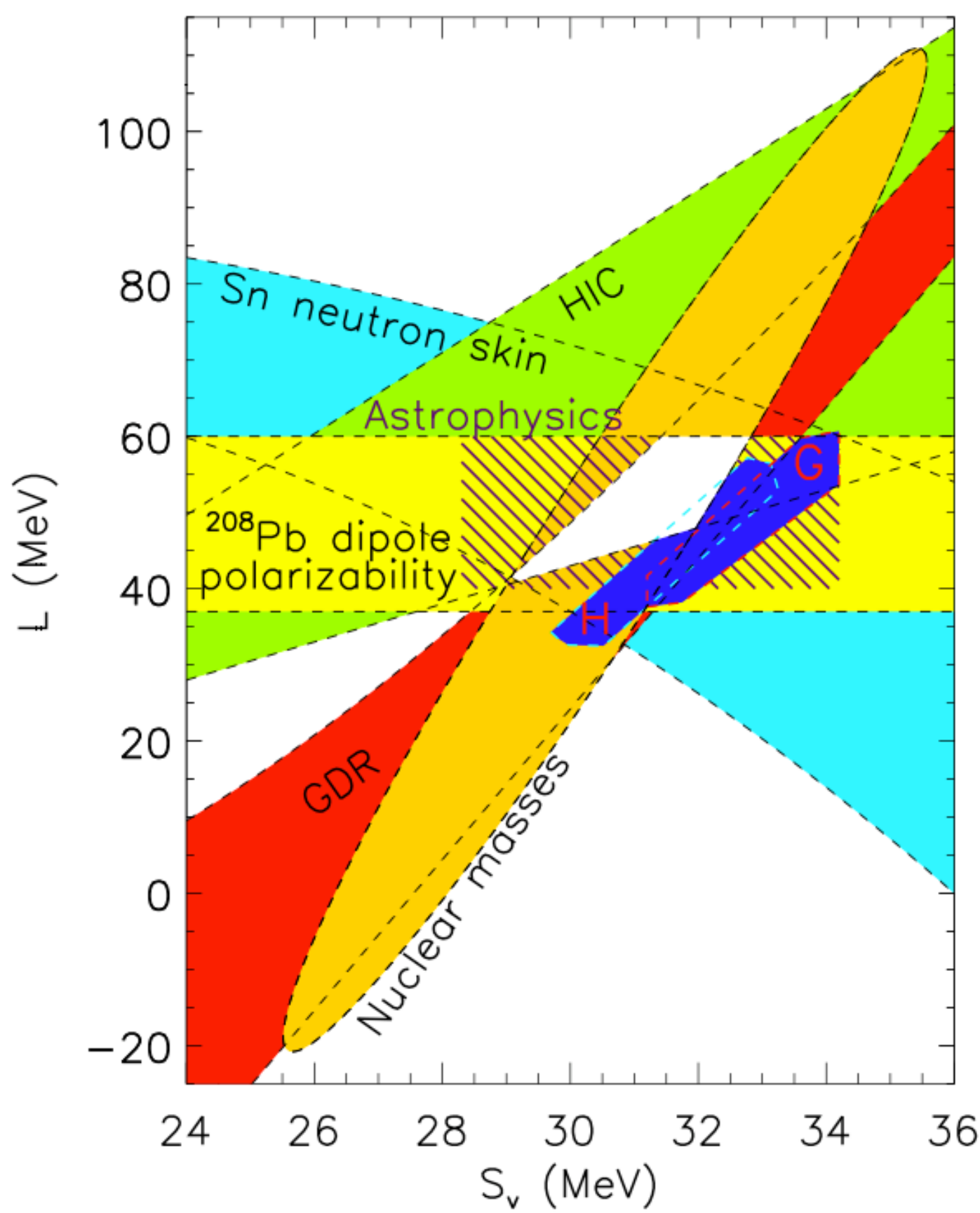}
\ece
\caption{\label{fig:polarizability}
Left panel: Correlation between $\Delta r_{np}$ and the electric dipole polarizability $\alpha_D$ in $^{208}$Pb established in Ref.~\cite{ReiNa:2010}. Right panel: astrophysical constraints on 
the (volume) symmetry energy $S(n_0)$  and its slope $L$~\cite{LaLi:2012}.}
\end{figure}

An alternative to obtain experimental information on $\Delta r_{np}$ of heavy nuclei and hence $L$ is through their static electric dipole polarizability, $\alpha_D$. The strong correlation between $\alpha_D$ and $\Delta r_{np}$ (left panel of Fig.~\ref{fig:polarizability}) has been established in mean-field models~\cite{ReiNa:2010,Pie:2010}. The nuclear dipole polarizability is defined through the frequency-dependent dipole strength function $S_D(\omega)=\sum_N|\!\bra{N}D\ket{0}\!|^2\delta(\omega-E_N)$ (or equivalently the photo-absorption cross section $\sigma_{abs}$) as  
\begin{equation}
 \alpha_D =
 \frac{8\pi e^2}{9} \int\!\! d\omega\,\frac{S_D(\omega)}{\omega}=
  \frac{1}{2 \pi^2 e^2} 
 \int\!\! d\omega\,\frac{\sigma_{abs}(\omega)}{\omega^2}\; , 
 \label{eq:pol}
 \end{equation}
where $\omega$ denotes the nuclear excitation (photon) energy.

Because of the inverse energy weighting, $\alpha_D$ sensitively depends on the $E1$ strength at low energies. A complete measurement of the dipole response $S_D(\omega)$ for $^{208}$Pb has recently been achieved through inelastic scattering of polarized protons at very forward angles at RCNP in 
Osaka~\cite{Tamii:2011}. The extracted dipole strength, being consistent with $\sigma_{abs}$ above the neutron-emission threshold, has also allowed to uniquely deduce the sub-threshold strength, which is crucial because of the inverse-energy weighting in Eq.~\ref{eq:pol}. 

Exploiting the tight correlation between $\alpha_D$ and $\Delta r_{np}$ 
(left panel of Fig.~\ref{fig:polarizability}), the measured value of $\alpha_D=20.1\pm 0.6$ fm$^3$/e$^2$~\cite{Tamii:2011} translates into a skin thickness  $\Delta r_{np}=0.156\pm 0.021$ fm, which is much more precise than the current PREX result. It enters prominently into the current contraints on the slope parameter $L$~\cite{LaLi:2012} (right panel of Fig.~\ref{fig:polarizability}).

\section{Inhomogeneous phases in the inner core}

There have been many speculations whether, in the inner core of a neutron star, the density is sufficiently large to induce a transition to a state in which quarks become deconfined. In this novel state a variety of new phases have been predicted, most prominenty chirally restored phases in which quarks loose their constituent mass as well as "color-superconducting" phases where quarks of different flavors appear in paired states. 

In most studies of chiral symmetry restoration in high-density quark matter it is tacitly assumed that the chiral order parameters of the various phases are uniform in space. On the other hand, they could be spatially modulated~\cite{Broniowski:2011}. In the meantime, several studies in QCD-like models such as the Nambu Jona-Lasinio model or the chiral quark-meson model have revealed that this is indeed a possibility. Most investigations on inhomogeneous chiral phases assume simplified forms for their spatial variation. A popular example is the 'chiral density wave' in which the (complex) chiral order parameter rotates uniformly in space. This is analogous to the so-called Fulde-Ferrel phases in (color-) superconductivity~\cite{Fulde:1964}.

\begin{figure}[tbh]
\includegraphics[width=8cm]{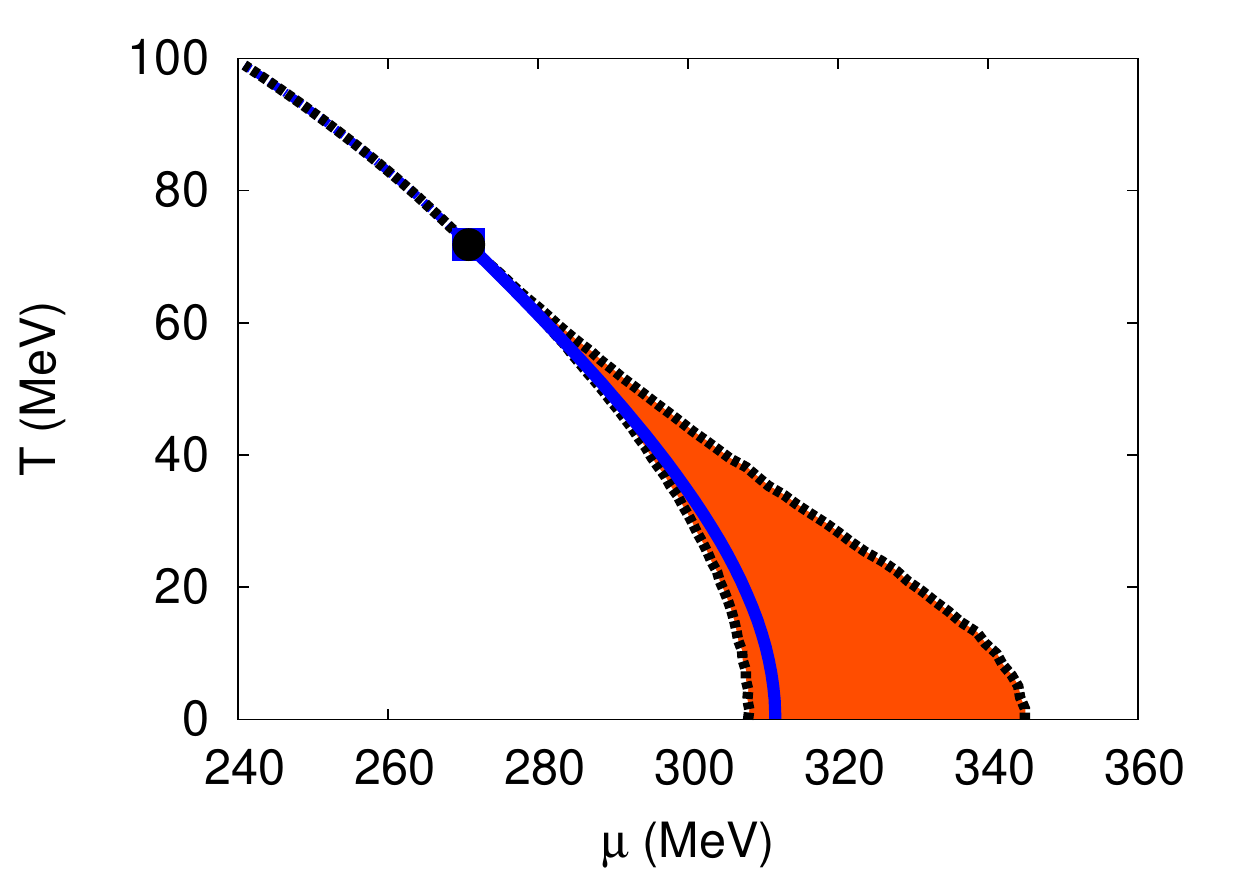}\includegraphics[width=8.0cm]{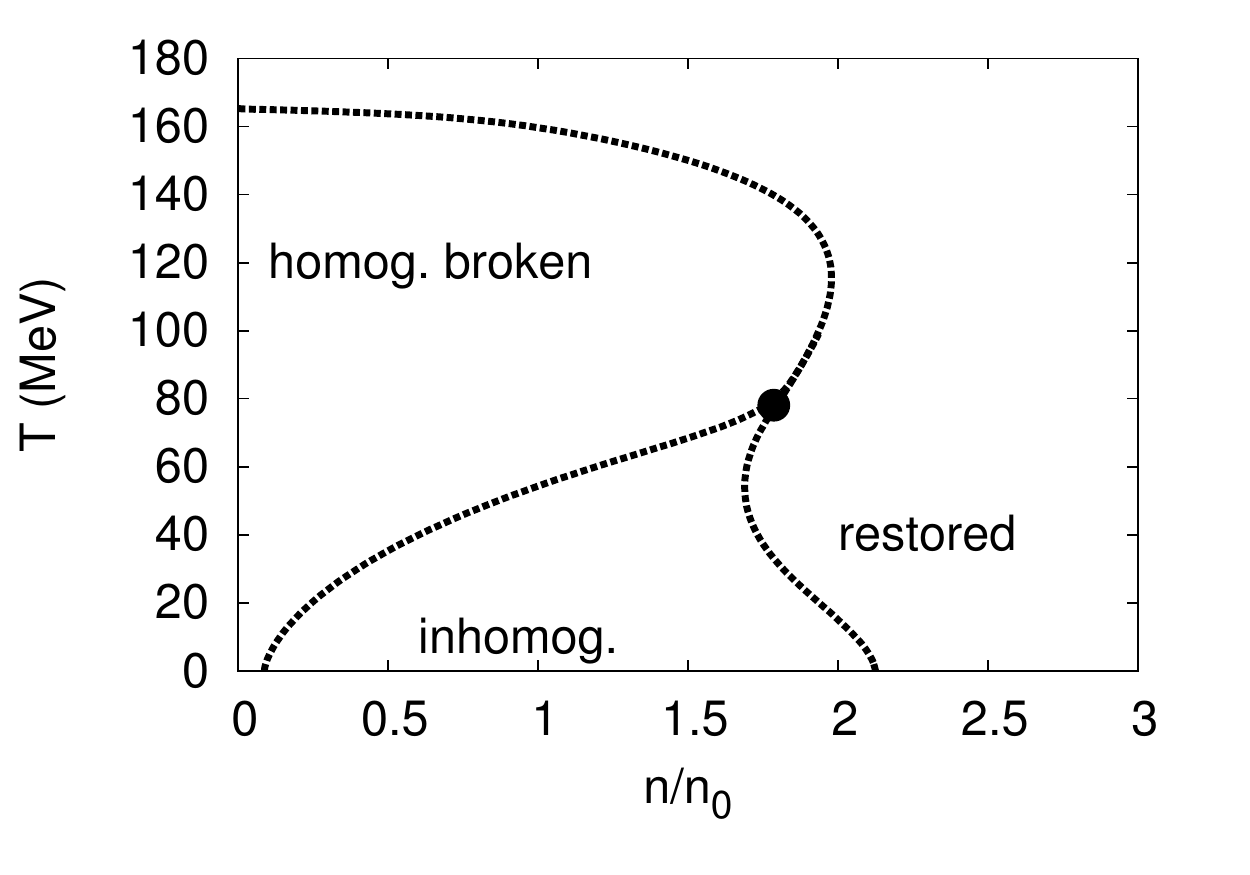}
\caption{\label{fig:inhomog}
Left panel: region of inhomogeneous 1d-chiral phases in the QCD phase diagram~\cite{Nickel:2009}. 
The solid blue line marks the phase boundary for a homogeneous first-order chiral transition. Right panel: chiral phase diagram in terms of number density $n$ rather than chemical potential~\cite{CNB:2010}.}
\end{figure}

More general spatial modulations are more difficult to obtain. By embedding exact one-dimensional 
solutions~\cite{Schnetz:2006} of QCD-like models in 3d-space~\cite{Nickel:2009} it has become possible, however, to find the energetically most favorable 1d-modulations (plates) as a function of temperature 
$T$ and chemical potential $\mu$. The resulting phase diagram, displayed in the left panel of 
Fig.~\ref{fig:inhomog}, indicates 
that the inhomogeneous phase covers the region where a first-order chiral transition would occur for a spatially 
homogeneous transition. The latter case features a line of first-oder transitions which ends in a critical point of second order, much like in a liquid-gas transition. Allowing for spatial inhomogeneities this chiral critical point disappears from the phase diagram, leaving only a 'Lifshitz point' in which three second-order lines meet~\cite{Nickel:2009,Nickel:2009prl}. Moving from low to high chemical potential, the spatial profile of the condensate changes gradually from a periodic kink solution to a sinusoidal modulation, whose amplitude decreases continuously until the chirally restored homogeneous phase is finally reached. The temperature dependence of the Lifschitz point and its density dependence (right panel of Fig.~\ref{fig:inhomog}) are rather insensitve to the assumptions of effective QCD-like models.

Limiting oneself  to one-dimensional structures is a strong assumption. Especially at lower temperatures, higher-dimensional modulations are to be expected. In fact, it has long been known that 1d-phases are unstable to thermal fluctuations and true long-range order cannot exist~\cite{Baym:1982}. It is therefore important to also investigate higher-dimensional modulations of the chiral order parameter.

\begin{figure}[tbh]
\begin{center}
\includegraphics[height=.4\textwidth,angle=270]{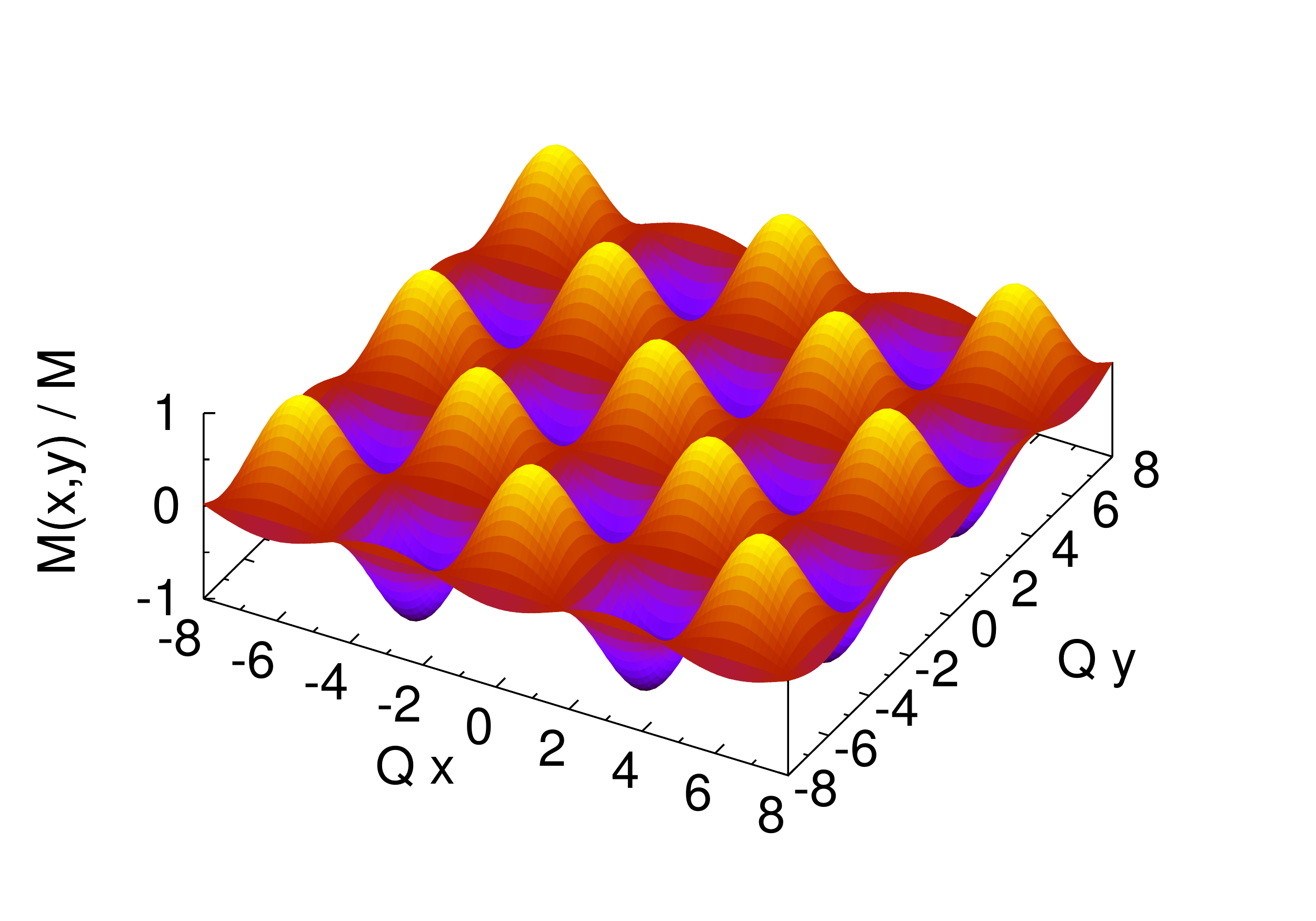}
\includegraphics[height=.4\textwidth,angle=270]{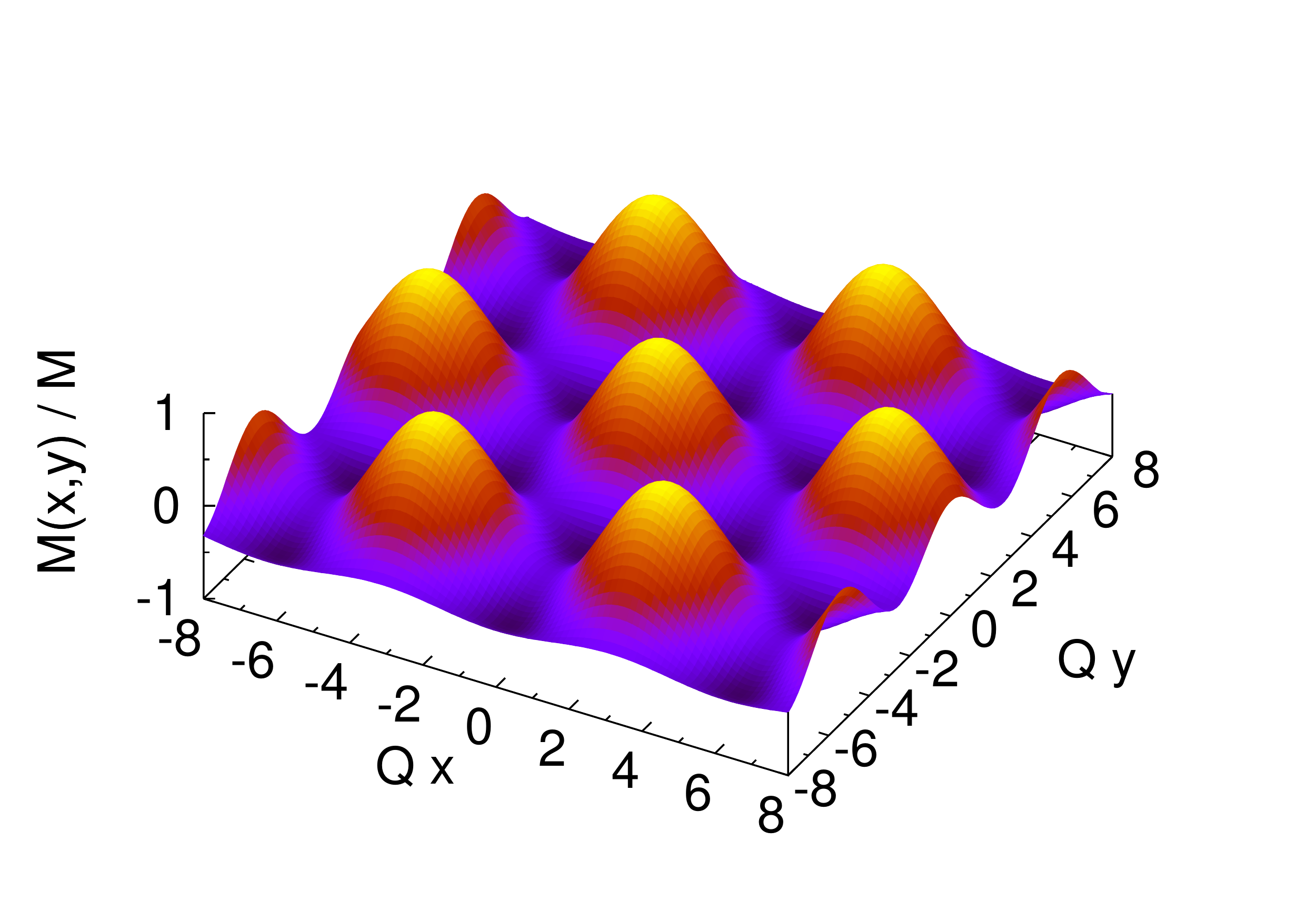}
\end{center}
\caption{Normalized mass functions $M(x,y)$ for two-dimensional spatial modulations of the chiral condensate.
Left panel: square lattice. Right panel: hexagonal lattice.} 
\label{fig:2dmods}
\end{figure}

The implemetation of general periodic structures turns out to be computationally demanding. Therefore, sofar, only 2d-structures have been looked at, assuming square- and hexagonal arrays of rod-like structures 
(Fig.~\ref{fig:2dmods}) with sinusoidal variation of chiral condensate (mass function)  in the $x$- and 
$y$-direction, i.e 
\bea
M(x,y)&=&M\cos(Qx)\cos(Qy)\quad{\rm square}
\nonumber\\
M(x,y)&=&\frac{M}{3}\left[2\cos (Qx)\cos\left(\frac{1}{\sqrt{3}}Qy\right)+\cos\left(\frac{2}{\sqrt{3}}Qy\right)\right]
\quad{\rm hexagonal}
\eea
where $M$ denotes the amplitude and $Q$ the wavevector.   

\begin{figure}[h!]
  \begin{center}
  \includegraphics[height=.45\textwidth,angle=270]{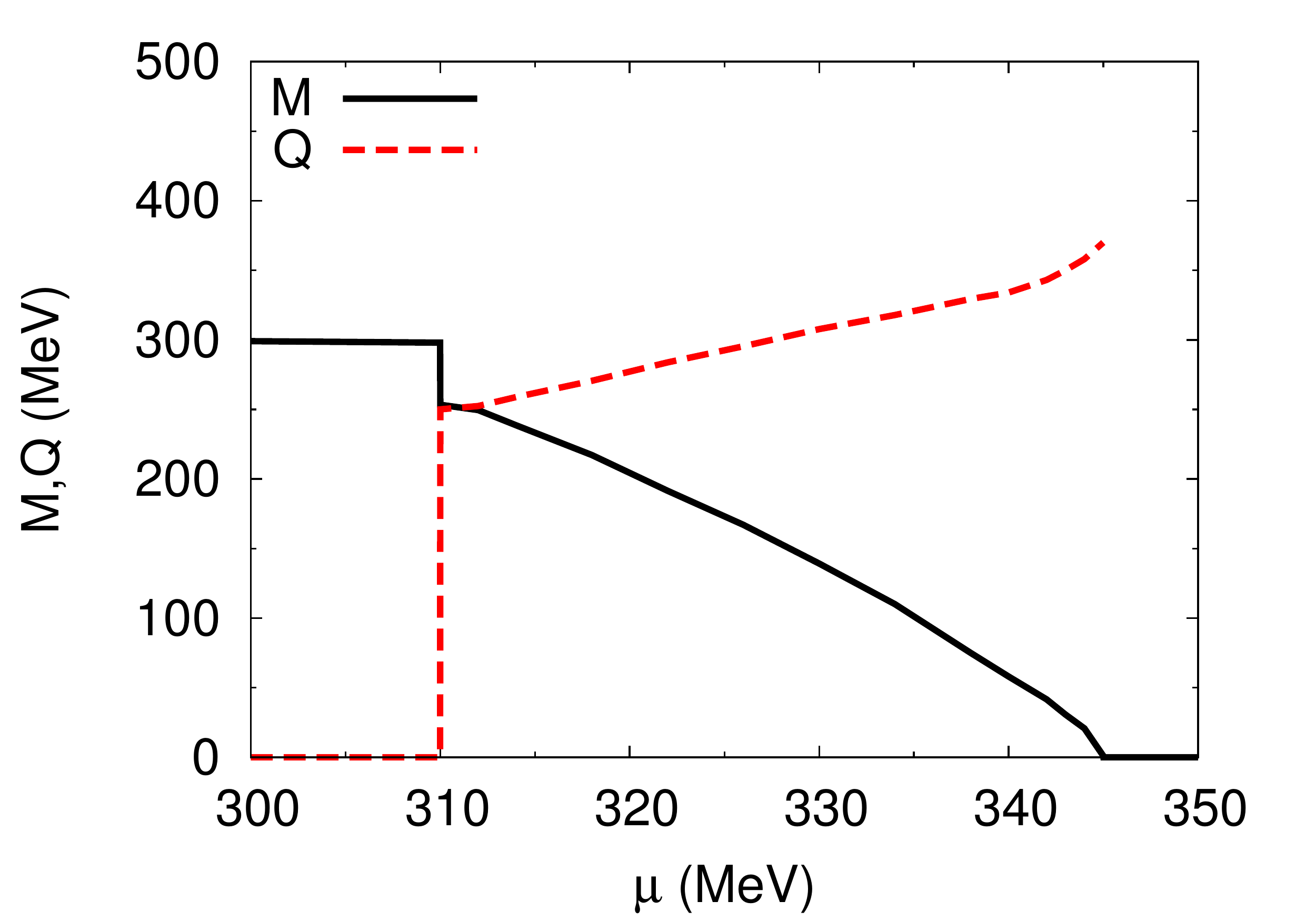}
  \includegraphics[height=.45\textwidth,angle=270]{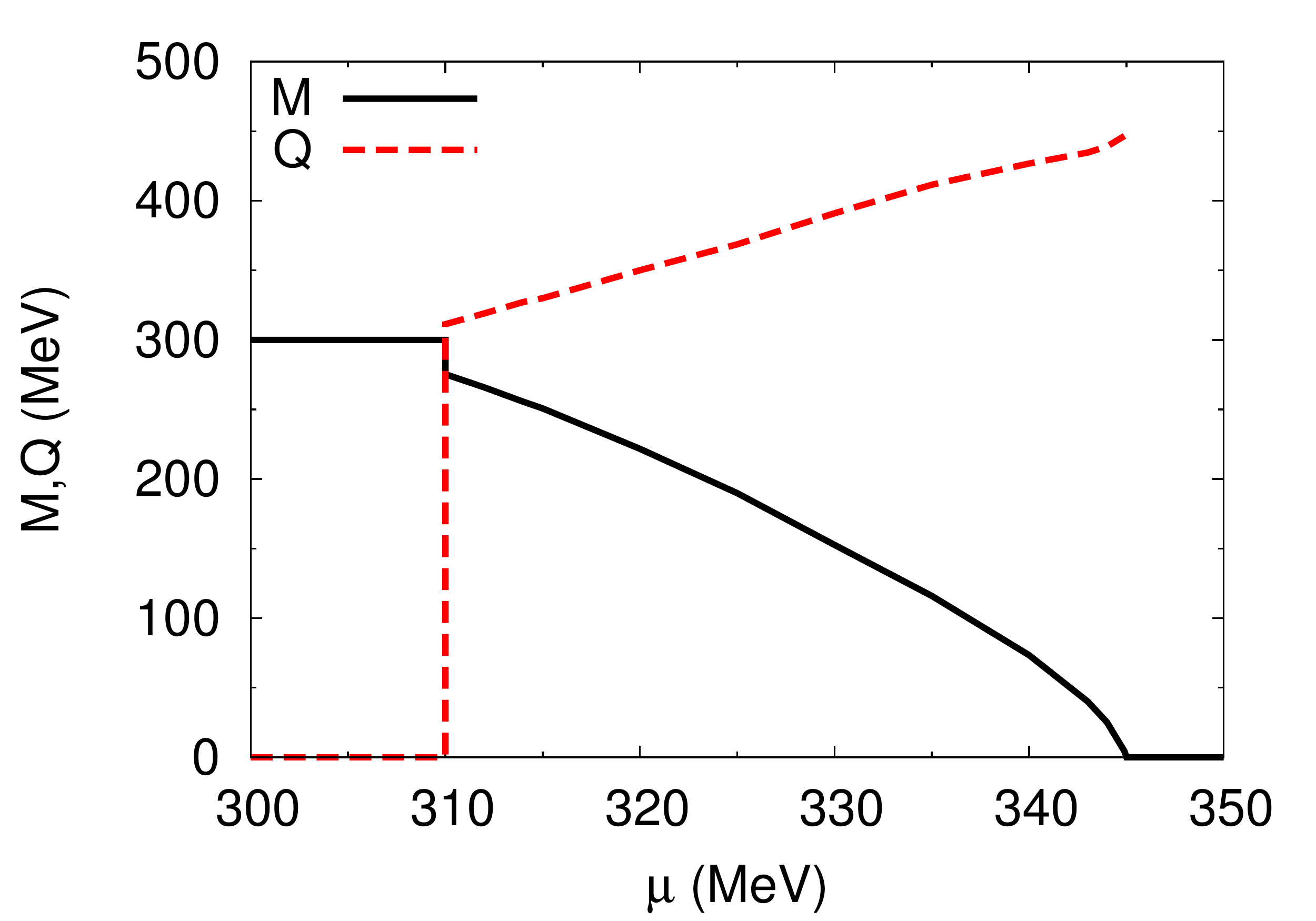}
  \end{center}
  \caption{Amplitude $M$ and wave number $Q$ at $T=0$ as functions of
  the chemical potential $\mu$ of 2d modulations the mass function~\cite{CB:2012}.
  Left panel: square lattice. 
  Right panel: hexagonal lattice. }
  \label{fig:tpmin2d}
\end{figure}

The results of the numerical minimization of the thermodynamic potential~\cite{CB:2012} are presented in Fig.~\ref{fig:tpmin2d}.
For both lattice geometries a sharp onset of the crystalline phase is observed
around $\mu \approx 310$ MeV followed by a smooth 
approach to the restored phase through a continuous decrease in amplitude and an increase in wavenumber $Q$. 

\begin{figure}[b!]
\begin{center}
\includegraphics[width=.55\textwidth]{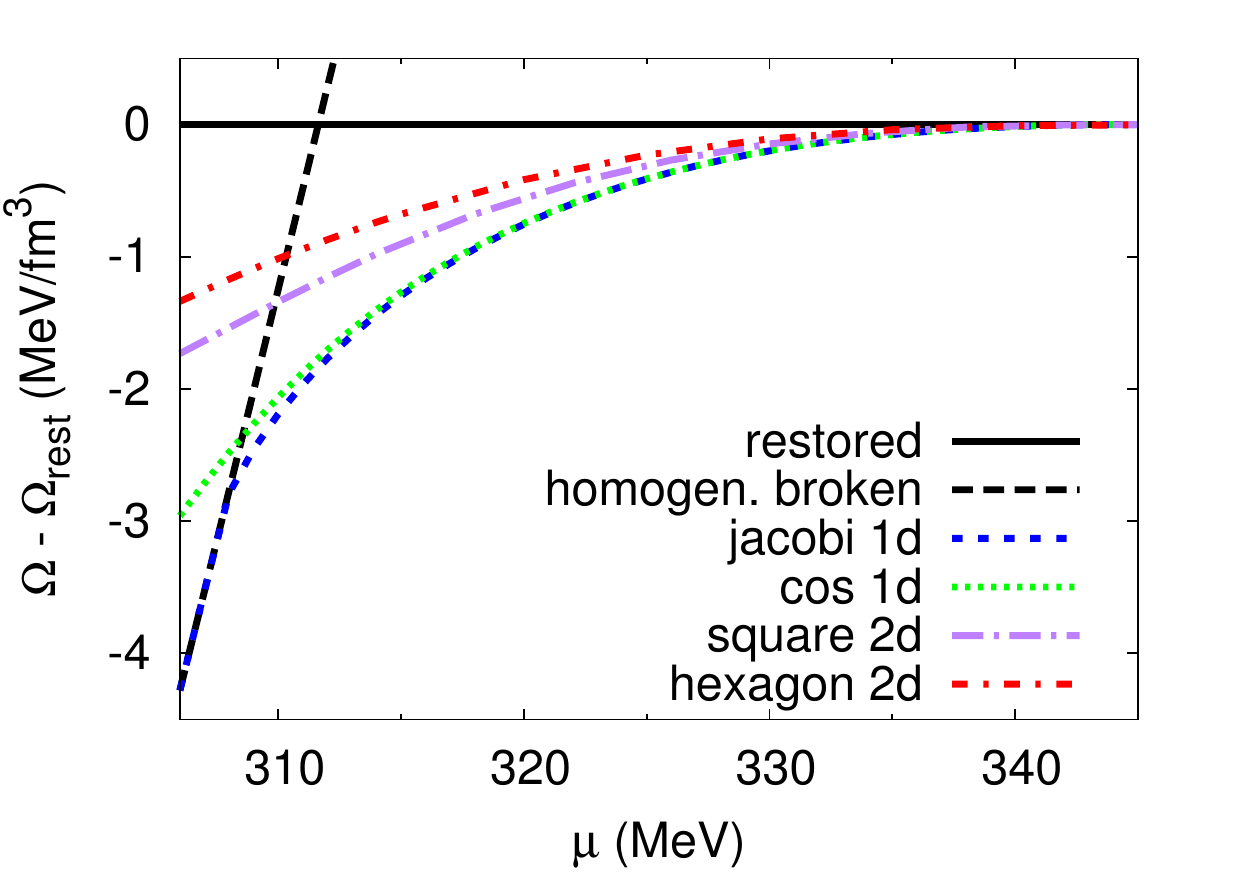}
\end{center}
\caption{
Thermodynamic potential relative to the restored phase for different 1d- and 2d modulations of the chiral condensate at $T=0$~\cite{CB:2012}.} 
\label{fig:cfrtp}
\end{figure}

To find the true ground state  one has to compare the free energies of the various phases with each other.
The results are displayed in Fig.~\ref{fig:cfrtp}.

One observes that the one-dimensional plate-like solutions lead to the biggest gain in free energy compared to all the other cases. In particular, the two-dimensional rod-like structures turn out to be energetically disfavored with respect to one-dimensional solutions throughout the whole inhomogeneous window.

\section{Summary and Conclusion}

In this contribution I have discussed two aspects in the physics of neutron stars. The first dealt with a recent measurement of the electric dipole polarizabitiy of $^{208}$Pb from which the neutron skin thickness can be determined rather precisely. This adds an important nuclear physics contraint to the symmetry energy and its derivative and hence the EoS of neutron matter. The second, more speculative, aspect focussed on possible inhomogeneous chiral phases in the inner core, provided deconfined quark matter would exist in this region. The energetically favored phases are acompanied by periodic density modulations which may have important implications for the transport properties of the inner core.

\bigskip
\noindent
{\it Acknowledgements:} I thank M. Buballa and S. Carignano for discussions on the second topic. This work has been supported in part by the Helmholtz Alliance EMMI and the Helmholtz International Center HICforFAIR.

\end{document}

%% file: econfmacros-csqcd3.tex
\def\beq{\begin{equation}}
\def\eeq#1{\label{#1}\end{equation}}
\def\eeqn{\end{equation}}

\def\beqa{\begin{eqnarray}}
\def\eeqa#1{\label{#1}\end{eqnarray}}
\def\eeqan{\end{eqnarray}}

\let\bar=\overbar

\def\bra#1{\left\langle{ #1} \right|}
\def\ket#1{\left| {#1} \right\rangle}

\def\Dslash{\not{\hbox{\kern-4pt $D$}}}
\def\dslash{\not{\hbox{\kern-2pt $\del$}}}

\def\msb{{\bar{\ssstyle M \kern -1pt S}}}

\usepackage{fancyhdr,graphicx}

%% file: CSQCD3.bbl
\begin{thebibliography}{99}

\bibitem{Demorest:2010} P. Demorest et al., Nature {\bf 467}, 1081 (2010).
\bibitem{Antoniadis:2013} J. Antoniadis et al., Science {\bf 340}, 1233232 (2013).
\bibitem{Wiki} http:$/\!/$en.wikipedia.org/wiki/file:neutron-star-structure.jpg
\bibitem{LaPr:2001} J. Lattimer and M. Prakash, Astrophys. J. {\bf 550}, 426 (2001). 
\bibitem{FuWo:2006} C. Fuchs and H.H. Wolter, Eur. J. Phys. {\bf A39}, 5 (2006).
\bibitem{ACVW:2011} X. Rocca-Maza et al., Phys. Rev. Lett. {\bf 106}, 252501 (2011b).
\bibitem{PiCe:2009} J. Piekarewicz and M. Centelles, Phys. Rev. {\bf C79}, 054311 (2009).
\bibitem{AAAAA:2012} S. Abrahamyan et al., Phys. Rev. Lett. {\bf 108}, 112502 (2012).
\bibitem{ReiNa:2010} P.-G. Reinhard and W. Nazarewicz, Phys. Rev. {\bf 81}, 051303(R) (2010).
\bibitem{Pie:2010} J. Piekarewicz, Phys. Rev. {\bf C83}, 034319 (2011).
\bibitem{Tamii:2011} A. Tamii et a., Phys. Rev. Lett. {\bf 107}, 062502 (2011).
\bibitem{LaLi:2012} J. Lattimer and Y. Lim, arXiv:1203.4286 [nuc-th].
\bibitem{Broniowski:2011} W. Broniowski,  Acta Phys. Pol. B Proc. Suppl. 5/3, 631 (2012).
\bibitem{Fulde:1964} P. Fulde and R.A. Ferrell, Phys. Rev. {\bf 135}, A550 (1964).
\bibitem{Schnetz:2006} O. Schnetz, M. Thies and K. Urlichs, Annals Phys. {\bf 321}, 2604 (2006).
\bibitem{Nickel:2009} D. Nickel, Phys. Rev. {\bf D80}, 074025 (2009).
\bibitem{CNB:2010} S. Carignano, D. Nickel and M. Buballa, Phys. Rev. {\bf D82}, 054009 (2010).  
\bibitem{Nickel:2009prl} D. Nickel, Phys. Rev. Lett. {\bf 103}, 072301 (2009).
\bibitem{Baym:1982} G. Baym, B. Friman and G. Grinstein, Nucl. Phys.  {\bf B210}, 193 (1982).  
 \bibitem{CB:2012}S. Carignano and M. Buballa, Phys. Rev. {\bf D86}, 074018 (2012). 
\end{thebibliography}
